\input{aipcheck}

\documentclass[
    ,final            
  ]
  {aipproc}

\layoutstyle{8x11single}

\begin{document}

\title{The critical end point through observables}


\author{G. Kozlov}{
  address={Joint Institute for Nuclear Research, Joliot-Curie st.6, Dubna, 141980, Russia}
}



\begin{abstract}
 We develop the model of  the critical phenomena of strongly interacting matter  at high temperatures and baryon densities.  The dual Yang-Mills theory with scalar degrees of freedom (the dilatons) is used. The dilatons are the consequence of a spontaneous breaking of a scale symmetry. The phase transitions are considered in systems where the field conjugate to the order parameter has the critical end mode. The critical end point (CEP) is a distinct singular feature existence of which is dictated by the chiral dynamics. The physical realization of CEP is via the influence quantum fluctuations of two-body Bose-Einstein correlations for observed particles to which the critical end mode couples. 
\end{abstract}

\maketitle


\section{Introduction}

The search of strong interacting matter at high temperatures $T$ and high baryon densities corresponding to the freezeout point is one of the issues  which will be in the focus of heavy ion machines, e.g., the collider NICA at JINR (Dubna) [1] and the  accelerator facility FAIR at GSI (Darmstadt) [2]. Both of the experimental facilities above mentioned are expected to be in operation soon after 2019. The freezeout point in particle physics is often referred to as the critical end point (CEP) of quantum chromodynamics (QCD).  The phase transitions in the proximity of CEP are associated with breaking of symmetry.
At high  $T$ and at finite baryon densities, where non-vanishing baryon chemical potentials $\mu$ are assumed, the  matter becomes weakly coupled and at the vicinity of CEP the color is no more confined, the chiral symmetry is restored. The CEP itself may be clarified through the search of its location on the $(\mu-T$) plane, the phase diagram, where each point  corresponds to a (metha)stable thermodynamic state characterized by $T$-dependent gauge-invariant functions. The properties of these states may be derived from the partition function
$$Z(T,\mu) = \sum _{i}\exp \left [-(E_{i} - \mu\, B_{i})/T\right ], $$
where $i$ labels the state with the energy $E_{i}$.
Thus, to realize the program of study the critical phenomena (e.g., to predict the CEP location)
 the high luminosity and high baryon density are required.

At large distances, QCD itself exhibits nonperturbative phenomena such as chiral symmetry breaking and confinement of color charges. The relation between these phenomena is not yet clarified in the frame of QCD, therefore the correlation or no one-to-one correspondence between phase transitions of chiral symmetry restoration and deconfinement in QCD at finite temperatures is an important issue. The analytical treatment of QCD at high $T$ is essential for the understanding of the phenomena above mentioned, however it is very difficult in the sense of its realization because of massless quarks in the theory.
One of the intrinsic approaches to analytical calculations can be seen through the scheme with topological defects which emerge in some effective models. The general properties of the field models containing the defects are related to that these defects exist only in the phase with spontaneously broken symmetry where the average expectations of the scalar fields do emerge. In the phase with nonbroken symmetry the solutions relevant to topological defects are absent.

The minimal model where the topological defects (e.g., strings)  arise is the Abelian Higgs-like model [3]. The key point is to reduce  the  $SU(N_c)$ gluodynamics to $[U(1)]^{N_c-1}$ dual Abelian scalar theory for $N_c$ color numbers. The breaking of the  gauge symmetry is realized through the Higgs-like mechanism.
We use the maximally Abelian gauge which suggests itself the special properties of QCD vacuum, such as Abelian dominance and the condensation of scalars [4], which provide the dual superconductor picture of the QCD vacuum [5]. In this vacuum the color-electric flux is squeezed into an almost one-dimensional object such as string due to the dual Meissner effect caused by scalar condensation. This is the dual analogy of the Abrikosov vortex in the classical superconductor supported by the Cooper pair condensation.

In lattice QCD by removing the scalar degrees of freedom (d.o.f.) the chiral symmetry breaking and the confinement are lost both simultaneously. This features the special role of a scalar field to both chiral symmetry breaking and confinement which lead to the fact that these two QCD nonperturbative phenomena are related to each other through the scalars (e.g., the monopoles or the dilatons). The essential point  is the description of the long-distance Yang-Mills (YM) physics in terms of  a dual gauge theory in which all particles become massive due to a dual scalar mechanism.

Below we  study the critical phenomena in strong interacting matter on a  qualitative  level of the non-perturbative dynamics  from the first principles. We deal in the framework of an effective Abelian model of $SU(3)$ gluodynamics which allows to describe the infra-red (IR) properties of the vacuum, and where an essential point is an appearance of scalar d.o.f, the dilatons, in the theory. The dilaton, the pseudo-Goldstone boson, is associated with spontaneous breaking of scale  symmetry of some four dimensional gauge theory. We develop an effective model in which the approximate scale invariance is manifest at high energies, but is spontaneously broken at a scale $f$ close to the QCD scale. The dilaton  is lighter than the other resonances which can have the masses $\sim 4\pi f$. We imagine that the entire structure of QCD is embedded in the conformal sector at high energies.

The physical approach to the QCD critical phenomena is supposed to be done  via the influence fluctuations of two-body Bose-Einstein (BE) correlations for observed particles to which the critical end mode couples. By studying BE correlations of identical particles (e.g., like-sign  charged particles of the same sort), one can estimate  the time and the spatial region (decoupling surface) over which the particles do not have the interactions. However, for an evolving system due to  heavy ion (or proton-proton) collisions, there is no a surface itself, since at each time there is a spread out surface because of (thermal) fluctuations in the final interactions and correlations. Thus, the shape of the surface evolves even in time and the particle emitted source is not approximately constant. The extended model of BE correlations at finite temperature approach to quantum field theory is carried out in [6], and this model can be applied to experimental data. 
One of the parameters of the model is the effective temperature of the particle emission source under a random   force  influence. 
 Based on the dual YM theory, we study the effect of CEP on two-particle BE correlation function $C_{2}$, in particular, on the fluctuations of the correlation strength  and the particle emission source size which both  can be extracted from experimental data. 
The quark-antiquark bound states are presented in terms of flux tubes the properties and correlations of which are under consideration. 
We suppose a strongly interacting medium is created in heavy ion  collisions which can exhibit the space momentum correlations and the collective behavior. 
One of the essential quantities that influences the particle freezeout is the formation time in the flux tube fragmentation. Note that some characteristics of $C_{2}$  may be singular at CEP.
The clear signature of the latter is non-monotonous behavior of $C_{2}$ characteristics which are rather sensitive to the proximity of CEP and they could be measured (or extracted) with the magnitude of the fluctuations strengths.

\section{Dual field correlator}
One of the features of CEP is a fluctuation measure related to some observables which may be visible through the fluctuations of some characteristic length as a derivative of the critical end mode. 
It means to find the fluctuations based on the probability distribution of an order parameter (gauge) field. 
The gauge-invariant two-point correlation function (TPCF) relevant to two gluon field strengths $F_{\mu\nu}$ at different (Euclidean) space-time points, connected by a Schwinger color string 
$$ U_{A}(x,y) = P\,\exp \left [i\,e\,\int_{y}^{x}\,dz^{\mu}\,A_{\mu} (z)\right ],$$
has the form [7]
\begin{equation}
\label{e1}
T_{\mu\nu\lambda\rho} (x) =\langle e{^2}\,F_{\mu\nu} (x)\,U_{A}(x,0)\,F_{\lambda\rho} (0)\,U_{A} (0,x)\rangle 
\end{equation}
with the following Lorentz decomposition
$$\left (\delta_{\mu\lambda}\delta_{\nu\rho} - \delta_{\mu\rho}\delta_{\nu\lambda}\right )e^{2} \,D_{1}(x^{2}) + \frac{1}{2}\left [\partial_{\mu}\left (x_{\lambda}\delta_{\nu\rho} - x_{\rho}\delta_{\nu\lambda}\right ) + 
\partial_{\nu}\left (x_{\rho}\delta_{\mu\lambda} - x_{\lambda}\delta_{\mu\rho}\right )\right ]e^{2}\,D_{2} (x^{2}),$$ 
where $F_{\mu\nu} = \partial_{\mu} A_{\nu} - \partial_{\nu} A_{\mu} - i\,e [A_{\mu},A_{\nu}]$;  
$A_{\mu} (x)= \sum_{a} A^{a}_{\mu}(x)\,t_{a}$, $A^{a}_{\mu}(x)$ are the YM fields; $t_{a}$ are the generators of the color gauge group in the fundamental representation having the standard commutation relations $[t_{a}, t_{b}] = i\,f_{abc}\,t_{c}$ and are normalized as $2\,tr~t_{a}\,t_{b} = \delta_{ab}$; $ e$ is the  coupling constant.
The leading tree level perturbative contribution is given by the form-factor $D_{2}$, while $D_{1}$ is different from zero and is dominated in IR region by decreasing behavior with the fall-off controlled by a finite mass parameter $M$.
The parametrization of the form 
\begin{equation}
\label{e3}
D_{1}(x^{2}) = A\,e^{-M\,\vert\vec x\vert} + \frac{a}{x^{4}}\,e^{-\tilde M\,\vert\vec x\vert} + ... 
\end{equation}
obeys TPCF (\ref{e1}) at short and large distances. The parameters $A, a, M, \tilde M$ in (\ref{e3}) are estimated through the lattice (cooled) data (see, e.g., [8] and the refs. therein). 
At large distances the first term in  (\ref{e3})  becomes important. 

It has been shown [8] that $D_{1}$  (\ref{e3}) is a nonvanishing function  if one assumes an effective dual approach to QCD where the main object is Abelian field strength $G_{\mu\nu} = \partial_{\mu} C_{\nu} - \partial_{\nu} C_{\mu}$, $C_{\mu}$ is the dual vector potential. 
The duality means that many features of color confinement in QCD could be understood if the continuum YM theory possesses some of properties of a magnetic superconductor (see, e.g., [9,10]). Therefore, the dual potentials $C^{a}_{\mu}$, instead of $A^{a}_{\mu}$, are the natural variables to be used in the confining regime.

In our model the duality means the assumption that physics of YM theory at large distances $r = \vert \vec x\vert$ with the strongly coupled gauge potential $A_{\mu}$ is the same as the  $r \rightarrow\infty$ physics of the dual theory describing the interactions of the potential $C_{\mu}$ weakly coupled to three scalar  fields $\phi_{i}\, (i=1,2,3)$ each in the adjoint representation of magnetic gauge group.
At large distances the dual to the dual field strength $\tilde G_{\mu\nu} =(1/2)\epsilon_{\mu\nu\alpha\beta}\,G_{\alpha\beta}$ behaves as the corresponding TPCF in QCD:
\begin{equation}
\label{e4}
\langle g^{2}\,\tilde G_{\mu\nu}(x)\,\tilde G_{\lambda\rho}(0)\rangle \sim
\langle e^{2}\,F_{\mu\nu}(x, x_{0})\,F_{\lambda\rho}(0,x_{0})\rangle, 
\end{equation}
where $g$ is the dual coupling constant, $x_{0}$ stands as an arbitrary reference point on the surface $S(\Gamma)$ (in the Wilson loop with the contour $\Gamma$) needed to surface ordering. 

\section{SCALE SYMMETRY BREAKING}
In four  dimensional gauge theories the field $\phi _{i}$ introduced in the previous section is associated with spontaneous breaking of (an approximate) scale symmetry. A light $CP$ -even scalar, a dilaton, may arise as a generic pseudo-Goldstone boson from the breaking of the conformal strong dynamics (see, e.g., [11]). 

We start with the partition function
\begin{equation}
\label{e5}
Z = \int  D\phi_{i}\,\exp \left [-\int_{0}^{\infty}\,d\tau\,\int\, d^{3} x\,L(\tau, \vec x)\right ],
\end{equation}
where the integral is taken over fields periodic in Euclidean time  $\tau$ with period $\beta$  in thermal (heatbath) equilibrium at temperature $T = \beta^{-1}$. In general, the Lagrangian density (LD) in (\ref{e5}) given in the operator form
\begin{equation}
\label{e6}
L(x) = \sum_{i} c_{i}(\mu)\, O_{i} (x)
\end{equation}
is renormalized at the scale $\mu$; $c_{i}(\mu)$ is running coupling,  and the operator $O_{i} (x)$ has the scaling dimension $d_{i}$. Under the scale transformations $x^{\mu}\rightarrow e^{\omega}\,x^{\mu}$ one has $O_{i} (x)\rightarrow e^{\omega\,d_{i}}\,O_{i} (e^{\omega}x)$, $\mu\rightarrow e^{-\omega}\mu$, $\omega$ is the parameter of a scale transformation. This gives for the  dilatation current $S^{\mu} = T^{\mu\nu}x_{\nu}$ [11]
$$\partial_{\mu} S^{\mu} = T_{\mu}^{\mu} = \sum_{i} \left [c_{i}(\mu) (d_{i} - 4) O_{i}(x) + \beta_{i} (c) \frac{\partial}{\partial c_{i}} L\right ],$$
where $T^{\mu\nu}$ is the energy-momentum tensor, $\beta_{i} (c) = \mu\partial c_{i} (\mu) /\partial\mu$ is the running $\beta$-function.  
The theory would be nearly scale invariant if $d_{i} = 4$ and $\beta (c)\rightarrow 0$. 
For $SU(N)$ theory with $N_{f}$ flavors in fundamental representation 
$$\beta (\alpha) = -\frac{b_{0}}{2\,\pi}\alpha^{2} -\frac{b_{1}}{(2\,\pi)^{2}}\alpha^{3} - ... ,$$
where $b_{0}$ and $b_{1}$ are well-known coefficients which can be found elsewhere.
The couplings $c_{i} (\mu)$ in (\ref{e6}) that, in general, are not scale invariant will obey the scale invariant sector by introducing the appropriate powers of the dilaton field to compensate the shift under the scale transformations. At energies below the conformal symmetry breaking scale $f$ the coupling 
$c_{i} (\mu)$ has to be replaced [11]
$$c_{i}(\mu)\rightarrow \left (\frac{\phi}{f}\right )^{4-d_{i}}\,c_{i}\left (\mu\frac{\phi}{f}\right),$$
where the incorporated flat direction (the dilaton field) $\phi$ transforms according to $\phi (x)\rightarrow e^{\omega}\phi (e^{\omega}x)$, and $f = \langle \phi\rangle$ is the order parameter for scale symmetry breaking, determined by the dynamics of the underlying strong sector.
For an approximate dilatation symmetry (small $\beta (\alpha)$)  the non-trivial IR fixed point  $\alpha^{\star} = -2\pi\,b_{0}/b_{1}$  may have to be strong enough (with small $N_{f}$), however, it exceeds a critical strength $\alpha_{c}$ (with respective $N^{c}_{f}$) for the spontaneous breaking of chiral symmetry that leads to appearance of confinement. The scale of confinement associated with the scale of chiral symmetry breaking is of the order of the QCD scale $\Lambda\sim O(0.5\, GeV)$ at which $\alpha (\mu)$ crosses $\alpha_{c}$. 
Thus, the breaking of chiral symmetry is triggered by the dynamics of nearly conformal sector. 

The theory with fixed $\alpha^{\star}$ now possesses an exact scale invariance in the limit of chiral symmetry. The scale invariance is spontaneously broken when the chiral symmetry is also broken spontaneously [12]. As a result, a massless scalar particles, the dilatons, can appear. Since the actual gauge coupling constant is running and the scale symmetry is explicitely broken, the dilaton associated with the condensate should appear as a pseudo-Goldstone boson with a mass of the order of  $\Lambda$.

\section{Dual model. Flux tubes}

In $SU(3)$ gluodynamics the effective LD of the dual model  is 
\begin{equation}
\label{e7}
L_{eff} = -\frac{1}{4}G_{\mu\nu}G^{\mu\nu} + \sum_{i=1}^{3}\left [\frac{1}{2} 
{\vert D_{\mu}^{(i)}\phi_{i}\vert}^{2} - \frac{1}{4}\lambda_{\phi} \left (\phi_{i}^{2} - \phi_{{0}_{i}}^{2}\right)^{2}\right ],
\end{equation}
where 
$$G_{\mu\nu} = \partial_{\mu} C_{\nu} - \partial_{\nu}C_{\mu} - i\,g [C_{\mu},C_{\nu}],\,\,
 D^{(i)}_{\mu}\phi_{i} = \partial_{\mu}\phi_{i} - ig [C_{\mu},\phi _{i}],\,\, C_{\mu}(x) = \sum_{a=1}^{8} C^{a}_{\mu} (x)\,t_{a}.$$

The coupling of the dilaton to dual field $C_{\mu}$ which becomes strong at the scale of chiral symmetry breaking is dictated by the appropriate group algebra relations.
In the $[U(1)]^{2}$ Higgs-like model, $D_{\mu}^{(i)} =\partial_{\mu} + i\,g\,\epsilon^{a}\,C_{\mu}^{a}$ is the covariant derivative acting on the scalar fields $\phi_{i}$, where $i=1,2,3$ and $a=3,8$. The $\epsilon$'s are the root vectors of the group $SU(3)$: $\vec\epsilon_{1} = (1,0)$,   $\vec\epsilon_{2} = (-1/2, -\sqrt {3}/2)$,  $\vec\epsilon_{3} = (-1/2, \sqrt {3}/2)$. The gauge fields $C_{\mu}^{a=3,8}$ are dual to the diagonal components of gluon fields   $A_{\mu}^{a=3,8}$. The scalar fields $\phi_{i}$  appear due to the compactness of the residual abelian gauge group $[U(1)]^{2}$ in the abelian projection $SU(3)\rightarrow [U(1)]^{2}$. The LD (\ref{e7}) is gauge invariant under  the shift $C_{\mu}^{a}\rightarrow C_{\mu}^{a} + \partial_{\mu} \alpha^{a}$, where $\alpha^{a=3,8}$ are the parameters of the gauge transformation. The condition $\sum_{i=1}^{3} arg\,\phi_{i} = 0$ reflects the restriction of the phases of $\phi_{i}$. The components of $G_{\mu\nu}$ define the magnetic field $\vec H$, $H^{k} = G^{0k}$, and 
the electric field $\vec\varepsilon $, $\varepsilon^{k} = (1/2)\epsilon_{klm} G^{lm}$. 

The dual Wilson loop [13]
$$U_{C}(x,y) =P\exp\left [i\,g\int_{y}^{x} dz^{\mu}\,C_{\mu} (z)\right ]$$
defines $C_{\mu} (x)$, where $U_{C}$ is invariant under gauge transformations 
$$C_{\mu} (x)\rightarrow\Omega^{-1}_{C}(x)\,C_{\mu}(x)\,\Omega_{C}(x) + \frac{i}{g}\,\Omega^{-1}_{C} (x)\,\partial_{\mu}\Omega_{C}(x),$$
and $\Omega_{C} (x)$ being an element of magnetic-color gauge group.
The scalar fields $\phi_{i}(x)$ with the order parameters 
$\langle \phi_{i} (x)\rangle =  \phi_{{0}_{i}}$, the vacuum expectation values (v.e.v.),  are associated with not individual particles but the subsidiary magnetically charged objects which cannot be observed experimentally. 
The color structure of $\phi_{{0}_{i}}$ is given by [14]
$$\phi_{{0}_{1}} = \frac {f}{\sqrt {2N}}\,J_{x}, \,\,\,\phi_{{0}_{2}} = \frac {f}{\sqrt {2N}}\,J_{y},\,\,\,\, \phi_{{0}_{3}} = \frac {f}{\sqrt {2N}}\,J_{z},$$
where $J_{j}$ are  three generators of the $N$ dimensional irreducible  representation of the three ($j$= $x$, $y$, $z$) dimensional rotation group corresponding to angular momentum $J = (N-1)/2$.
In the confinement phase the dual  gauge symmetry is broken due to dual scalar mechanism, and all the particles become massive. The quanta of $C_{\mu}$ acquires a mass $m\sim gf$ via the dual Higgs-like mechanism, hence the dual theory is weakly coupled at distances $r > 1/mass$, where the denominator being either the mass of dual gauge quanta or the mass $m_{\phi}\sim\sqrt {2\lambda_{\phi}}\,f$
of the scalar field with the coupling constant $\lambda_{\phi}$ (see (\ref{e7})). 
Using the scheme with the partially conserved dilatation current  [15], where the dilaton mass $m_{\phi}$ is scaled by $\Lambda$  
$$ m_{\phi} \simeq \sqrt {1 - \frac{N_{f}}{N^{c}_{f}}}\,\Lambda, \,\,\,\, N_{f} \leq N^{c}_{f},$$
the v.e.v. $f$ can be  estimated through the decay constant $f_{\pi}$ of the $\pi$-meson (in QCD $f_{\pi}\simeq 0.3 \Lambda$) 
$$f\simeq \frac{1}{0.3\,\sqrt {2\lambda_{\phi}}} \sqrt {1 - \frac{N_{f}}{N^{c}_{f}}}\, f_{\pi}. $$

The interaction between two charges is due to color magnetic current $J_{\mu} (x) \sim \partial^{\nu} G_{\mu\nu} (x)$ in the scalar (dilaton) condensate, where $G_{\mu\nu} = \partial_{\mu}\,C_{\nu} - \partial_{\nu}\,C_{\mu} +G^{s}_{\mu\nu}$. The $G^{s}_{\mu\nu}$ is the Dirac string tensor representing a moving line from the charge $-g_{m}$ to the charge $+g_{m}$
$$G^{s}_{\mu\nu} (x) = g\,\epsilon_{\mu\nu\alpha\beta}\,\int_{0}^{1}d\tau\int_{0}^{1}d\sigma\,\frac{d y_{\alpha}}{d\sigma}\frac{d y_{\beta}}{d\tau}\,\delta^{4} [x - y (\tau,\sigma)],$$
where $y_{\mu} (\tau,\sigma)$ is a world sheet of a surface $S (\Gamma)$ swept by the Dirac string connecting $-g_{m}$ and $+g_{m}$. The fixed contour $\Gamma$ on the $S(\Gamma)$ depends on the charge source trajectories $z_{\mu}$: $\Gamma [z_{{1}_{\mu}} = y_{\mu}(\tau,\sigma =1),  z_{{2}_{\mu}} = y_{\mu}(\tau,\sigma =0)]$. The divergence of dual to Dirac string tensor is the current carried by a charge $g_{m}$ moving along the path $\Gamma$: $\partial^{\beta}\tilde G^{s}_{\alpha\beta} (x) \sim g\int dz_{\alpha}\,\delta ^{4} (x-z)$. The quantization condition $e\cdot g = 2\,\pi$ guarantees that the Dirac string will not be observable when the dual gauge symmetry is  not broken.

The excitations above the (classical) vacuum in the effective theory are flux tubes connecting a quark-antiquark pair in which $Z_{N}$ electric flux is confined to the narrow tubes of a radius $\sim m^{-1}$, at whose center the scalar condensate vanishes.
The probability distribution related to the ensemble of systems containing a single flux tube with $N(R)$ number of configurations of the flux tube of the length $R$ is [16] 
\begin{equation}
\label{e8}
Z_{flux}(\beta, R, m) = \sum_{\beta}\sum_{R}\,N(R)\,\exp \left [-\beta\,E (m,R)\right ]\,D (\vert \vec x\vert, \beta; M),
\end{equation}
where $E (m, R)$ is the effective energy of the peace of the isolating string-like tube 
$$E (m, R)\sim m^{2}\,R [a + b\,\ln (\tilde\mu\,R)],$$
$a$ and $b$ are known constants, $\tilde\mu$ is some massive parameter. The mass $m$  of  $C_{\mu}$-field  develops an infinite fluctuation  length $\xi\sim m^{-1}$ in the proximity of  CEP with  the critical temperature $T_{c} = \beta_{c}^{-1}$. Note, that $m^{2}(\beta)\sim g^{2}(\beta)\,\delta^{(2)}(0)$, where $\delta^{(2)}(0)$ is the inverse cross-section of the flux tube, which could be expressed in terms of the string radius $r_{s}$ as follows: $\delta^{(2)}(0)\sim 1/(\pi\,r^{2}_{s})$.

The function $D$ in (\ref{e8}) is associated with the scalar part of TPCF (\ref{e1}) in the form of large distances exponential fall-off of correlator of gauge-invariant operators $ O$ 
\begin{equation}
\label{e9}
\langle O(\tau,\vec x)\,O(\tau, 0)\rangle \sim const \,{\vert \vec x\vert}^{c}\, D(\vert \vec x\vert, \beta; M), 
\end{equation}
$$D(\vert \vec x\vert, \beta; M) = \exp \left [- M(\beta)\,\vert\vec x\vert\right ], $$
where the existence of $D$ at $T > T_{c}$ is admitted; $c$ in (\ref{e9})  is the constant depending on the choice of the operator $O(\tau, \vec x)$. Within the dual conformity (\ref{e4}) we assume that $M^{-1}(\beta)$ is the measure of the screening effect of color electric field. In $SU(N)$ theory with $N = 2,3$ at hight $T$ and zero chemical potential with $N_{f}$ massless quark flavors, $M(\beta)$ is [17]
$$M(\beta) = M^{LO}(\beta) + N\,\alpha\,T\,\ln\left [\frac{M^{LO}(\beta)}{4\,\pi\,\alpha\,T}\right ] + 4\,\pi\,\alpha\,T\,y_{n/p} (N) + O(\alpha^{3/2} T,\alpha^{2} T), $$
where 
$$M^{LO}(\beta) = \sqrt {4\,\pi\,\alpha\left (\frac{N}{3} + \frac{N_{f}}{6}\right )}\,T,$$
$y_{n/p}  = 1.58\pm 0.20$ for $SU(2)$ and  $y_{n/p}  = 2.46\pm 0.15$ for $SU(3)$.
At high $T$, one has $\alpha << 1$, and the correlator  (\ref{e9}) is ($c= -4$)
\begin{equation}
\label{e10}
\langle O(\tau,\vec x) O(\tau, 0)\rangle \sim L^{-4}_{W} - \frac{T}{V_{W}}\sigma_{0}\xi^{2} \left [\frac{1}{\xi}\sqrt {\frac{8 \pi}{\sigma_{0}}} -N \ln \left (\xi\sqrt {2 \pi\sigma_{0}}\right ) + 4 \pi y_{n/p}+ ...\right ].
\end{equation} 
We find that  the effective theory in terms of nonperturbative TPCF describes the fluctuations at distances  $g\,\xi/\sqrt{\pi} < \vert\vec x\vert < M^{-1}$ up to CEP; $\sigma_{0} \simeq (3/4) \alpha (\beta)\,m^{2}(\beta)$ [16]. In order to get the result (\ref{e10}), $\vert \vec x\vert$ is replaced by the spatial Wilson loop  $L_{W}$, which has an area law behaviour below and above $T_{c}$ ($V_{W}$ in (\ref{e10}) is the volume subject to $L_{W}$).

TPCF (\ref{e10}) has a singular behavior at small distances, however it disappears at large fluctuations  ($\xi\rightarrow\infty$) when CEP approaches.
In the scheme with the flux tubes, $\xi$ is the penetration depth of a color-electric field (or, approximately,  the radius of the flux tube), while the inverse dilaton mass $l = m^{-1}_{\phi}$ stands for the coherent length of the scalar field (condensate). 
The formation time of the flux tube is $\tau_{flux} = \sqrt {4/(3\alpha)}\xi$ which tends to infinity at CEP.  For $SU(3)$, $m\simeq 2\sqrt {\sigma_{0}}$ [18], and the lattice simulations yield $T_{c}\simeq 0.65 \sqrt {\sigma_{0}}$ [19]. Therefore, the effective theory should also be applicable in the deconfined phase with  temperatures $T_c < T < 3 T_{c}$. 
 

We consider $N(R)$   in (\ref{e8}) in the discrete space of a dilaton condensate, where 
$N(R) = V\,l^{-3}\,exp[s(R/l)]$, and the flux tubes with the entropy density $s$ lie along the links of a 3-dimensional cubic lattice of a volume $V$ with the lattice size $l << R$. Then the partition function $Z_{flux}$ (\ref{e8}) for the flux tube is
\begin{equation}
\label{e11}
Z_{flux} (\beta, R, m) = \frac{V}{l^{3}}\sum_{R} \exp (-E_{flux}\,\beta), 
\end{equation}
where the energy of the flux tube  is $E_{flux} (\beta) =\sigma_{eff} (\beta)\,R$ and
\begin{equation}
\label{e12}
\sigma_{eff} (\beta) = \sigma_{0} - \frac{s}{l\,\beta} + \frac{\vert\vec x\vert}{R\,\beta}\,M(\beta).
\end{equation}
The sum of first two terms in (\ref{e12}) is  the order parameter of phase transition when the chiral symmetry is restored at $T = T_{c}$;
$s = E_{tot}/T$, and the total energy $E_{tot}$ is identified with the mass of a quark-antiquark bound state $m_{q\bar q}$. 

The interactions between flux tubes are defined by the scalar and gauge boson fields profiles. 
 The ratio $k_{GL} = \xi/l$ as the Ginzburg-Landau-like parameter defines the properties of the dual superconductor QCD vacuum. The attracted forces can appear between two (parallel) flux tubes (of the same type) if $k_{GL} < 1$ (type-I vacuum), otherwise the flux tubes repel each other in the vacuum where $k_{GL} >1$ (type-II vacuum), and deconfinement is characterized by $k_{GL}\rightarrow \infty$. 

 The phase transition, if occurred, can be seen  through the singularity once partition function is calculated. The Wilson loop string tension $\sigma_{0} - s/(l\,\beta)$ in (\ref{e12}) disappears at $T = T_c$.
In the non-perturbative regime the quark-antiquark bound states still exist at $T > T_{c}$ as Goldstone bosons with (deconfined) string tension $\sigma _{D} \sim \alpha\,T^{2}$. This gives an evidence of the magnetic component of deconfined phase where (thermal) dilatons evaporate from magnetic condensate at low $T$. Since dilatons remain massive up to the singular (end) point, there is a first order phase transition.  
The critical temperature $T_c$ can be found from the condition $\sigma_{0} = s\,m_{\phi}\,T_{c}$.
Keeping in mind that  a) $m = (4/3)\,\alpha^{-1}\,m_{q\bar q}$ in the border between the vacua of the types I and II ($k_{GL} =1$, $ m_{\phi} = m$) and b) for lattice $T_c = 0.65\,\sqrt {\sigma_{0}}$ [19] one has $ T_{c}\simeq$ 172 MeV for $\pi$ - mesons ($\alpha = 0.37$), while the value of baryon (chemical) potential $\mu{_c}$ is compared to  $m_{q\bar q}$.
The singularity of $Z_{flux}$ (\ref{e11}) may arise when the vacuum criterium obeys the following condition
\begin{equation}
\label{e13}
k_{GL} \geq \frac{3}{4}\frac{\alpha (\beta)}{\xi\,m_{\bar q q}}\left [1 + \frac{4}{3}\frac{\xi^{2}}{\alpha(\beta)\,\beta} M(\beta)\frac{L_W}{R}\right ],
\end{equation} 
where the first term in the  r.h.s of (\ref{e13}) stands for the type-I vacuum, while the second term enters the type-II vacuum and tends to infinity at large enough fluctuations $\xi$ accompanying $\alpha(\beta)\rightarrow 0$ as $\beta\rightarrow\beta_{c}$.

\section{Particle correlations and observables}

The flux tubes above mentioned are the concrete physical realization of the causal random (stochastic) processes. The latter are the open systems which are restricted and can interact with (infinite) environment. The open system accompanied by the heat bath is the complete system where any dynamical fluctuations might be occurred through the calculation of TPWF or the string tensions either in confinement or deconfinement phases.

On the other hand, the chiral symmetry breaking is characterized by the order parameter which is a chiral condensate $\langle q\bar q\rangle$ given by zero-eigenvalue density $\rho (0)$ 
\begin{equation}
\label{e14}
\langle q\bar q\rangle = - \lim_{mass\rightarrow 0}\,\lim_{V_{phys}\rightarrow\infty} \pi\,\rho (0),\,\,\, \rho(\kappa) = \frac{1}{V_{phys}} \sum _{n}\langle \delta (\kappa - \kappa_{n})\rangle
\end{equation} 
in the physical space-time volume $V_{phys}$, where $\kappa _{n}$ is the Dirac eigenvalue of the eigenstate $\vert n\rangle$, and the Dirac operator $\hat D =\gamma_{\mu}\,D^{\mu}$ acts as $\hat D\vert n\rangle = i\,\kappa _{n}\vert n\rangle $, $\langle m\vert n\rangle = \delta_{m,n}$.
 
The fluctuations of some modes in the vicinity of CEP can not be measured in experiments. These fluctuations can affect the  observables either in direct channel or in  indirect reactions. One of the examples is the  BE correlation phenomena where the strength of the correlation between  particles may have the influence fluctuations.

The two-particle BE correlation function 
\begin{equation}
\label{e15}
C_{2} (q, P) = \left\langle\frac{\int dx_{1}\,dx_{2}\,S (x_{1},p_{1})\,S (x_{2},p_{2}){\vert \Phi\vert}^{2}}
{\int dx_{1}\,S (x_{1},p_{1})\,\int dx_{2}\,S (x_{2},p_{2})}\right \rangle_{\beta}
\end{equation} 
in the sense of complete system  is given by the Wigner-like thermalized phase-space density $S(x,P)$ (an emission function) in the emitting system containing two particles with four-momenta $p_{1}$ and $p_{2}$, and it can be viewed as the probability that a particle with the average momentum $P = (p_{1} + p_{2})/2$ is emitted from the space-time point $x$ in the collision region; $q = p_{1} - p_{2}$; $\Phi$ is the two-particle state function. The index $\beta$ in the brackets (\ref{e15}) reflects the thermal bath influence.

At finite temperatures (\ref{e15})  has the form [6]
\begin{equation}
\label{e16}
C_{2} (q,\beta) \simeq \eta (n)\left\{1+ \lambda(\beta) e^{-q^{2}L^{2}_{st}} \left [ 1 + \lambda_{1}(\beta) e^{+q^{2}L^{2}_{st}/2}\right ]\right \},
\end{equation} 
where for $n$ particle multiplicity $\eta (n) = \langle n(n-1)\rangle/\langle n\rangle^{2}$; $L_{st} $ is the measure of the space overlap between two identical particles (flux tubes) affected by stochastic forces in the vacuum characterized by $k_{GL}$. 
The strengths $\lambda (\beta)$ and  $\lambda_{1} (\beta)$ in (\ref{e16}) related to coherent or chaotic character of emission effects in the particle source are [16]
\begin{equation}
\label{e17}
\lambda (\beta) =\frac{\gamma (\omega,\beta)}{ (1 + \nu)^{2}},\,\, \lambda_{1} (\beta) = \frac{2\,\nu}{\sqrt {\gamma(\omega,\beta)}},
\end{equation}
and the stochastic forces influence is given by  $\nu = \nu (\omega,\beta) < \infty $. 
The function $\gamma (\omega,\beta)$ in (\ref{e17}) calls for quantum thermal properties of the particle source:
$$ \gamma (\omega,\beta) = \frac{\hat n^{2} (\bar\omega)}{\hat n(\omega)\,\hat n(\omega^{\prime})}, \,\,\, \hat n(\omega) = \frac{1}{e^{(\omega - \mu)\beta} -1}, \,\,\, \bar\omega = \frac{\omega + \omega^{\prime}}{2}, $$
$\omega$ is the energy of the particle with momentum $p = (\omega,\vec p)$ in thermal bath with statistical equilibrium.

The strength  $\nu $ in (\ref{e17}) is an effective number of hadrons with the mass $m_{h}$ in the plane phase space with the size $L_{st}$ having the mean mass $\bar m_{h} (\omega, \beta) = m_{h}\,\hat n(\omega,\beta)$: 
\begin{equation}
\label{e18}
\nu (m_{h}, \beta) \sim \frac{ \hat n(\omega,\beta)}{\bar m^{2}_{h} (\omega, \beta)\, L^{2}_{st}}.
\end{equation}
Obviously, $\nu (m_{h},\beta)\rightarrow 0$ as $\hat n(\omega,\beta) \rightarrow\infty$ at high enough $T\rightarrow T_{c}$ and at fixed $\mu = \mu_{c}$.  
Note that in the quantum-mechanical scheme (see, e.g., [20]) $\lambda$ is a positive $c$-number restricted by 1, and no $\lambda_{1}$ appears in the formula for $C_{2}$.


{\it \bf   Results.}

  {\it 1. Dip-effect.}
At low temperatures, $T < \sqrt {m^{2}_{h} + \mu^{2}}$, in the case of light hadrons the function $C_{2}$ can approach even below unity (dip-effect) in the small region of $q$ with $\langle n\rangle \sim O(10)$, 
where the (correlated) system defined by the evolving stochastic scale 
\begin{equation}
\label{e19}
L_{st}\simeq \left [\frac{(2\,\pi)^{3/2}\,e^{\sqrt {m^{2}_{h} + \mu^{2}}\,\beta}}{3\,\nu (n)\,k^{2}_{T}\,(m^{2}_{h} + \mu^{2})^{3/4}\,T^{3/2}\,\left (1 + \frac{15}{8}\frac{T}{\sqrt {m^{2}_{h} + \mu^{2}}}\right )}\right ]^{1/5}
\end{equation} 
is disturbed by external force strength $\nu (n)$. 
In formula (\ref{e19}), the pair average transverse momentum $k_{T}$ is defined as half of the absolute vector sum of the two transverse momenta, $k_{T} = \vert \vec p_{T_{1}} + \vec p_{T_{2}}\vert /2$; $\vert \vec p_{T_{i}}\vert = \sqrt {\vec p^{2}_{x_{i}} + \vec p^{2}_{y_{i}}}$. The dependence of BE correlation signal on $k_{T}$ has been observed at the  SPS [21], at the Tevatron [22] and at RHIC [23]. 
 
The dip-effect,  unlike the monotonic shape of $C_{2}$ with $q$, where  $C_{2} (q\rightarrow\infty) =1$, has already been observed (as the "anticorrelation effect") in CMS experiment at the LHC [24] at $q\simeq [0.5 - 1.5]$ GeV depending on charged multiplicity $n$. The same qualitative behavior has been found earlier in L3 experiment at LEP [25]. At more higher temperatures  
the dip-effect disappears as $\langle n\rangle >> 1$, and at the CEP the $C_{2}$ function does not deviate from 1. 

 {\it 2. Evolving size.} The size of the particle emission source in terms of $L_{st}$ is strongly dependent on $\nu (n)$, $k_{T}$, $m_{h}$ and $T$. We have shown in (\ref{e19}) that at low temperatures 
$L_{st}$ decreases with $k_{T}$ and increases with $n$ ($\nu (n)\rightarrow 0$), where 
\begin{equation}
\label{e20}
\nu (n) = \frac{2 - \tilde C_{2}(0) + \sqrt {2 - \tilde C_{2}(0) }}{\tilde C_{2}(0) -1},\,\, \,\,\,\tilde C_{2}(0) =\frac{C_{2}(q=0)}{\eta (n)} 
\end{equation}
with $\langle n\rangle \geq 1 + C_{2}(0) /2$. From theoretical point of view $C_{2} (0)$ can not exceed the value of 2.

At higher temperatures there is a nontrivial singular behavior of $L_{st}$ with $\nu (n)$ 
\begin{equation}
\label{e21}
L_{st} \sim \left [\nu (n)\,k_{T}^{2}\,T^{3}\right ]^{-1/5}, 
\end{equation} 
where 
$$\nu(n)\sim \frac{1}{n\,k^{2}_{GL}} O\left (\frac{m_{\phi}^{2}}{m^{2}_{h}}\right )\rightarrow \frac{1}{n\,k^{2}_{GL}}~~~ as~~ \beta\rightarrow \beta_{c},$$  
and no the dependence of  $\mu$ and  $m_{h}$ are found in (\ref{e21}). Obviously, $L_{st}\rightarrow\infty$ as $\nu (n)\rightarrow 0$ with $n\rightarrow\infty $. The singular behavior of $L_{st}$ is consistent with chiral symmetry breaking (see (\ref{e14})), where the chiral condensate $\langle q\bar q\rangle $ is defined in infinite physical volume.
The CMS experiment [24] has found out that the effective emission radius increases with charged-particle multiplicity in the events with colliding energies from  $\sqrt {s}$ = 0.9 TeV to 7 TeV, and it decreases smoothly with $k_{T}$. The latter effect is more sufficient in events with large particle multiplicities. 
It is clearly seeing from (\ref{e21}) that CEP is a well-defined singularity on $L_{st}$ with respect to $\nu (n)$.
Thus, at the temperatures close to $T_{c}$ one can expect the strengthening of the expansion of the particle emission size with particle  multiplicity.  

 {\it 3. Correlation strength.} We emphasize that
the correlation strength $\lambda (\beta)$ decreases with $k_{T}$ (see (\ref{e17}) - (\ref{e19})).  The result of smooth decreasing of $\lambda$ with $k_{T}$ with slight increasing of the values of $\lambda$
at small $n$ was  demostrated by CMS (see Fig.3 in [24]). 
Actually, in the phase of deconfinement $\lambda (\beta_{c}) = \gamma (\beta_{c}) \simeq 1$, $\lambda_{1}(\beta_{c}) = 0$
as $T\geq T_{c}$  with the fluctuation length $\xi\rightarrow\infty$ or $L_{st} \rightarrow\infty$.
These observations are clearly seen through (\ref{e18}) - (\ref{e21}).  


\section{Conclusions}

To conclude, we studied the critical phenomena related to color confinement/deconfinement and the chiral symmetry breaking/restoration starting from first principles of strong interacting theory. The approach is based on $SU(3)$ YM theory described by an effective model coupling the dual   gauge potentials $C_{\mu}$ to 3 adjoint scalar fields, the dilatons. These couplings generate color magnetic currents which, via a dual Meissner effect, confine the electric flux to narrow tubes connecting a quark-antiquark pair. The dilaton associated with the condensate may be difficult  to detect as it has the vacuum quantum numbers.
Relevant to first order phase transition of $SU(3)$ the singular end point is characterized by the critical temperature $ T_{c}\simeq$ 172 MeV for $\pi$ - mesons, while the baryon chemical potential is about the mass of a hadron that corresponds to BE condensation.
The importance of a role of the dilaton in our consideration can be seeing immediately when one can find that confinement and chiral symmetry breaking are simultaneously lost if the dilaton are removing from the vacuum. This indicates that confinement and chiral symmetry breaking are strongly correlated to each other
via the  dilaton.

The effect of CEP on the particle emission size $L_{st}$ and the correlation strength $\lambda$ of fluctuations of BE correlations as the experimental observables in heavy-ion collisions is explored. The characteristic signature is non-monotonous  and even singular behavior  of $L_{st}$, which is sensitive to the proximity of CEP and is measured by the magnitude of fluctuation length $\xi$.  The $C_{2}$ being non-monotonous function of  $\sqrt {s}$, $T$, $k_{T}$, $\nu(n)$ has a trivial behavior defined by $\eta (n)\sim O(1)$ as the CEP is approached and then passed.

Finally, the following phenomenological quantities can indicate the occurrence of phase transition (CEP is approached):\\
 - $L_{st}$ blows up as $T\rightarrow T_{c}$ (due to $\nu (n)\rightarrow 0$, $m_{h}\rightarrow 0$); the singularity of $L_{st}$  is evident;\\
 - large enough fluctuation length $\xi$ ($\nu (n)\rightarrow 0$, chiral symmery is restored);\\
 - the function $C_{2}$  being non-monotonous one of $k_{T}$ at low $T$ does not deviate from 1 as $T \geq T_{c}$;\\
 - the correlation strengths $\lambda =1$ (fully chaotic source) and $\lambda_{1} = 0$ as $T \geq T_{c}$;\\
 - the allocation of CEP satisfies  $C_{2}\sim \lambda \sim 1$ relevant to $(\mu - T)$ phase diagram.





\bibliographystyle{aipproc}   

\bibliography{sample}

\IfFileExists{\jobname.bbl}{}
 {\typeout{}
  \typeout{******************************************}
  \typeout{** Please run "bibtex \jobname" to optain}
  \typeout{** the bibliography and then re-run LaTeX}
  \typeout{** twice to fix the references!}
  \typeout{******************************************}
  \typeout{}
 }



\end{document}